\documentstyle[pra,aps,epsf,preprint]{revtex}

\newcommand{\ket}[1]{\mathop{\left| #1 \right\rangle}\nolimits}
\newcommand{\bra}[1]{\mathop{\left\langle #1 \right|}\nolimits}
\newcommand{\braket}[2]{\mathop{\left\langle #1 \left| #2 \right.
\right\rangle}\nolimits}
\newcommand{\avr}[1]{\mathop{\left\langle #1\right\rangle}\nolimits}

\begin{document}
\draft

\tighten  

\title{Coherent information analysis of quantum channels in
simple quantum systems}
\author{B.\ A.\ Grishanin and
V.\ N.\ Zadkov\thanks{zadkov@comsim1.ilc.msu.su}}
\address{International Laser Center and Department of Physics\\
M.\ V.\ Lomonosov Moscow State University, 119899 Moscow, Russia}
\date{March 15, 2000}
\maketitle

\begin{abstract}
The coherent information concept is used to analyze a variety of
simple quantum systems. Coherent information was calculated for
the information decay in a two-level atom in the presence of an
external resonant field, for the information exchange between two
coupled two-level atoms, and for the information transfer from a
two-level atom to another atom and to a photon field. The coherent
information is shown to be equal to zero for all full-measurement
procedures, but it completely retains its original value for
quantum duplication. Transmission of information from one open
subsystem to another one in the entire closed system is analyzed
to learn quantum information about the forbidden atomic transition
via a dipole active transition of the same atom. It is argued that
coherent information can be used effectively to quantify the
information channels in physical systems where quantum coherence
plays an important role.
\end{abstract}

\narrowtext

\section{Introduction}
\label{sec:intro}

The concept of noisy quantum channel may be used in many
information-carrying applications, such as quantum communication,
quantum cryptography, and quantum computers \cite{qc-book}.
Shannon's theory of information
\cite{shannon48,shannon49,gallagher68,grover91} is a purely
classical one and cannot be applied to quantum mechanical systems.
Therefore, much recent work has been done on quantum analogues of
the Shannon theory
\cite{schumacher96,schum96,bennett96,lloyd96,barnum98,preskill}.
The {\em coherent information} introduced in
\cite{schum96,lloyd96} is suggested to be analogous to the concept
of {\em mutual information} in classical information theory. It is
defined by
\begin{equation}
\label{Ic}
I_c=S_{\rm out}-S_e,
\end{equation}

\noindent where $S_{\rm out}$ is the entropy of the information
channel output and $S_e$ is the {\em entropy exchange}
\cite{schumacher96,lloyd96} taken from the channel reservoir. If
$S_{\rm out}-S_e>0$, then, expressed in {\em qubits}, $I_c$ describes a
binary logarithm of the Hilbert space dimension, all states of
which are transmitted with the probability $p=1$ in the limit of
infinitely large ergodic ensembles. Otherwise, we set $I_c=0$.

The validity of the coherent information concept was proved in
\cite{lloyd96,barnum98}, and it was  used successfully for
quantifying the resources needed to perform physical tasks.
Coherent information is expected to be as universal as its
classical analogue, Shannon information, and it characterizes a
quantum information channel regardless of the nature of both
quantum information and quantum noise. In contrast to Shannon
information in classical physics, however, coherent information is
expected to play a more essential role in quantum physics. The
capacity of information channels in classical physics can be
estimated, in most cases, even without relying on any information
theory, at least within an order of magnitude. This, however, is
not feasible in quantum physics and the coherent information
concept, or a similar concept, must be used to quantify the
information capacity of the channel. An analysis of the quantum
information potentially available in physical systems is
especially important for planning experiments in new fields of
physics, such as quantum computations, quantum communications, and
quantum cryptography\cite{qc-book,preskill}, where the coherent
information of the quantum channel determines its potential
efficiency.

\begin{figure}
\begin{center}
\epsfxsize=5.cm\epsfclipon\leavevmode\epsffile{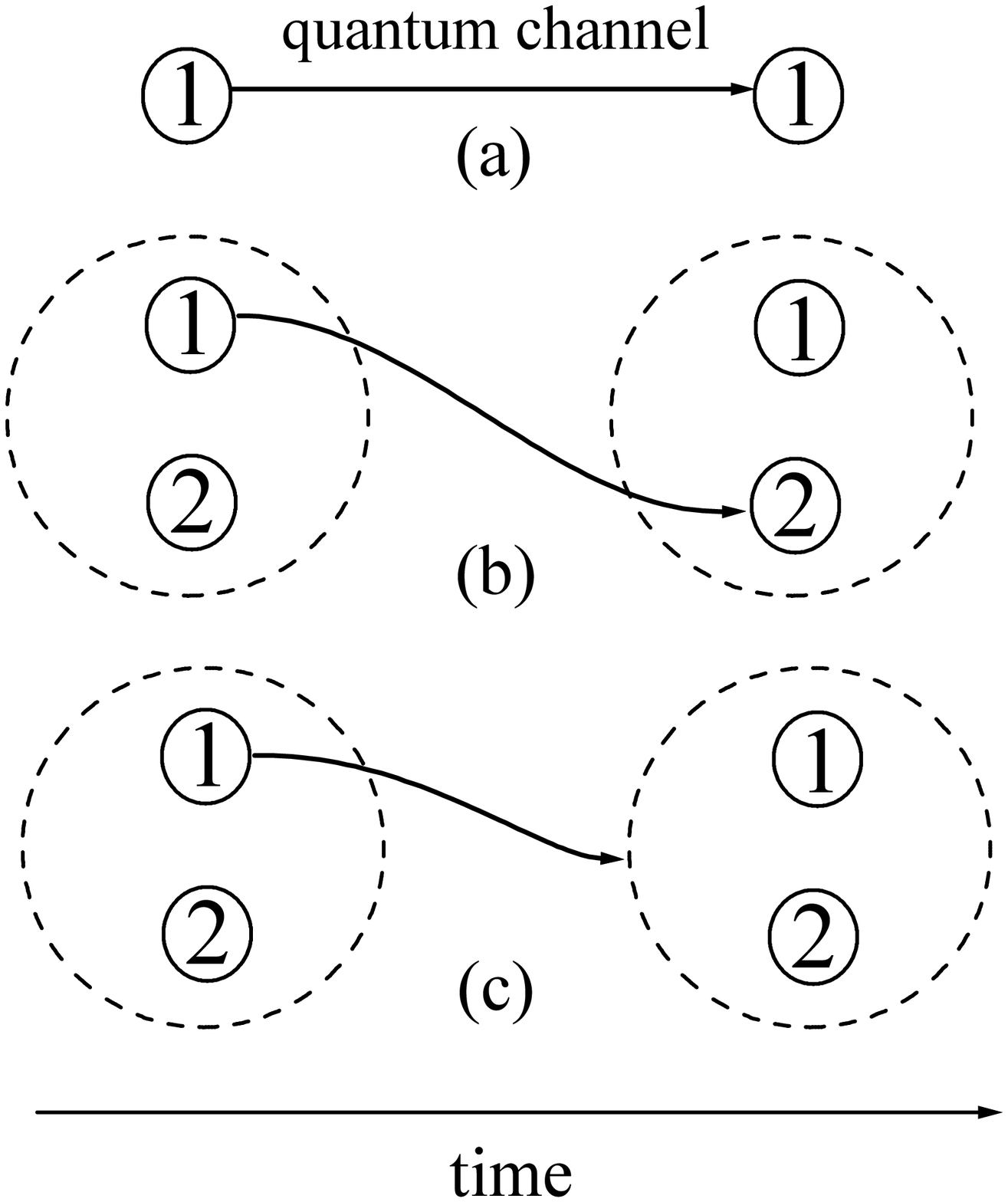}
\end{center}
\caption{Classification of possible quantum channels connecting two
quantum systems. $1\to1$, information is transmitted from the initial
state of the system to its final state (a); $1\to2$, information is
transmitted from subsystem 1 of the system (1+2) to subsystem 2
of the system (b); $1\to(1+2)$, information is transmitted from
subsystem 1 of the system (1+2) to the whole system (1+2) (c).}
\label{fig1}
\end{figure}

In this paper, we apply the coherent information concept to an
analysis of the quantum information exchange between two systems,
which in general case may have essentially different Hilbert
spaces. For this purpose, we must specify the noisy quantum
information channel and its corresponding superoperator $\cal S$,
which transforms the initial state of the first system into the
final state of another system. A classification scheme for
possible quantum channels connecting two quantum systems is shown
in Fig.\ \ref{fig1} \cite{note}. In addition to the two-time
channels shown in the figure, we consider also their one-time
analogues. Two-time quantum channels are widely used in quantum
communications and measurements, whereas one-time quantum channels
are appropriate for quantum computing and quantum teleportation.

The paper is organized as follows. In section
\ref{sec:definitions}, we explain key definitions and review
superoperator representation technique, which is used throughout
the paper. In the following sections we consider a variety of
quantum channels that correspond to the classification scheme
shown in Fig.\ \ref{fig1}. Section \ref{sec:onequbit} discusses
the coherent information transfer between quantum states of a
two-level atom (TLA) in a resonant laser field at two time
instants (Fig.\ \ref{fig1}a). The same type of quantum channel
($1\to1$) can be considered for a system that contains two (or
more) subsystems. This case is analyzed in section
\ref{sec:qubit-intra}, using a spinless model of the hydrogen atom
as an example. Coherent information transfer between two different
quantum systems is considered in section \ref{sec:twoqs}. The
analysis includes coherent information transfer between (i) two
unitary coupled TLAs (Fig.\ \ref{fig1}b), (ii) two TLAs coupled
via the measuring procedure (Fig.\ \ref{fig1}b), (iii) an
arbitrary system and its duplication (Fig.\ \ref{fig1}c), (iv) a
TLA in the free space photon field (Fig.\ \ref{fig1}b), and (v)
two TLAs via the free space photon field (Fig.\ \ref{fig1}b).
Finally, section \ref{sec:conclusions} concludes with a summary of
our results.

\section{Key definitions and calculation technique}
\label{sec:definitions}

\subsection{Notations and superoperator representation technique}
\label{subsec:notations}

This subsection explains key notations and briefly reviews the symbolic
superoperator representation technique \cite{gKE79}, which is especially
convenient for mathematical treatment of coherent information transmission
through a noisy quantum channel.

The most general symbolic representation of a superoperator is defined by
the expression
\begin{equation}
\label{genS}
{\cal S}=\sum\hat s_{kl}\bra{k}\odot\ket{l},
\end{equation}

\noindent where the substitution symbol $\odot$ must be replaced
by the transforming operator variable and $\bra{k}$ is an arbitrarily
chosen vector basis in Hilbert space $H$, to which the transformed
operators are applied. In order to correctly apply this
transformation to a density matrix, operators $\hat s_{kl}$ must
obey the positivity condition for the block operator $\hat S=(\hat
s_{kl})$ \cite{positivity} and orthonormalization condition
\begin{equation}
\label{snorm}
{\rm Tr}\,\hat s_{kl}=\delta_{kl},
\end{equation}

\noindent which provides normalization for all normalized operators
$\hat\rho$ with ${\rm Tr}\,\hat\rho=1$.

Using symbolic representation (\ref{genS}), one can easily
represent the production of superoperators ${\cal S}_1$, ${\cal
S}_2$, which constitutes a symbolic representation of the
superoperator algebra. For $\hat s_{kl}=\ket{k}\bra{l}$ it results
to the identity superoperator, ${\cal I}$, and for $\hat s_{kl}
=\ket{k}\bra{k} \delta_{kl}$---to the quantum reduction
superoperator ${\cal R} =\sum \ket{k} \bra{k} \odot
\ket{k}\bra{k}$. The case of $\hat s_{kl}= \delta_{kl}$ represents
the trace superoperator ${\rm Tr}\odot,$ which is a linear
functional in the density matrix space. The correspondence between
the matrix representation $S=(S_{mn})$ of the superoperator ${\cal
S}$ in orthonormalized operator basis $\hat e_k$ and its symbolic
representation (\ref{genS}) is given by
\begin{equation}
\label{skl}
\hat s_{kl}= {\cal S}(\ket{k}\bra{l})=
\sum\limits_{mn}S_{mn} \bra{l}\hat e_n\ket{k}\hat e_m
\end{equation}

\noindent and can be easily checked by substituting it in Eq.\
(\ref{genS}) and comparing with the standard definition of matrix
elements ${\cal S}\hat e_n=\sum_m S_{mn}\hat e_m$.

\subsection{The calculation of coherent information}
\label{subsec:general}

The entropy exchange in Eq.\ (\ref{Ic}) for the coherent
information is defined as
\begin{equation}
\label{Se}
S_e=S(\hat\rho_\alpha),\quad S(\hat\rho)=-{\rm
Tr}\,\hat\rho\log_2\hat\rho,
\end{equation}

\noindent where the joint input-output density matrix $\hat\rho_\alpha$
is given, in accordance with \cite{lloyd96,conjugation}, by
\begin{equation}
\label{inout}
\hat\rho_\alpha=\sum\limits_{ij}\,{\cal S} (\ket{\rho_i} \bra{\rho_j})
\otimes \ket{\bar{\rho}_i}\bra{\bar{\rho}_j}.
\end{equation}

\noindent Here $\ket{\rho_i}=\hat\rho_{\rm in}^{1/4}\ket{i}$ are
the transformed eigenvectors of the input density matrix
$\hat\rho_{\rm in}=\sum p_i\ket{i}\bra{i}$, bar symbol stands for
complex conjugation, and ${\cal S}$ is the channel input-output
superoperator, so that the output density matrix $\hat\rho_{\rm
out}={\cal S} \hat \rho_{\rm in}$. Using superoperator
representation (\ref{genS}) within the above defined eigen basis
$\ket{i}$, the density matrix (\ref{inout}) takes the form:
\begin{equation}
\label{inouts}
\hat\rho_\alpha=\sum\limits_{ij}(p_i^{}p_j^{})^{1/4}\,\hat s_{ij}^{}
\otimes \ket{\bar{\rho}_i}\bra{\bar{\rho}_j},
\end{equation}

\noindent where operators $\hat s_{ij}$ represent the states of
the output. Both the input and output marginal density matrices
are given by the trace over the corresponding complementary
system: $\hat \rho_{\rm out} ={\rm Tr}_{\rm in} \hat\rho_\alpha$,
$\hat{\bar{\rho}}_{\rm in}={\rm Tr}_{\rm out} \hat\rho_\alpha$. Finally,
the coherent information (\ref{Ic}) can be calculated, keeping in
mind that $S_{\rm out}=S(\hat\rho_{\rm out})$.

\subsubsection{Two-time coherent information for two quantum systems}
\label{subsubsec:twosystem}

For the coherent information transfer between two quantum systems
through the quantum channels shown in Figs \ref{fig1}b,c ($1\to2$ or
$1\to(1+2)$), the initial joint density matrix must be taken in the
product form $\hat\rho_{1+2}= \hat\rho_{\rm in}\otimes\hat\rho_2$, where
$\hat\rho_{\rm in}=\hat\rho_1$ and $\hat\rho_2$ are the initial
marginal density matrices, the first one being an input. For the
$1\to2$ quantum channel, the output is the state of the second system,
since a transformation on these two systems is made and a certain
amount of information is transmitted into the second system from the
initial state of the first one.

The dynamical evolution of the joint (1+2) system is given by a
superoperator ${\cal S}_{1+2}$ and the corresponding channel
transformation superoperator, which converts $\hat\rho_{\rm
out}={\cal S}\hat\rho_{\rm in}$, can be written as
$$
{\cal S}={\rm Tr}_1\,{\cal S}_{1+2}(\odot\otimes\hat\rho_2),
$$

\noindent where the trace is taken over the final state of the first
system. The transformation is described in terms of Eq.\ (\ref{genS})
for the joint system as
\begin{equation}
\label{skl12}
{\cal S}=\sum\limits_{k\kappa\;l\lambda}\sum\limits_n \bra{n}
\hat s_{k\kappa,l\lambda} \ket{n}\bra{\kappa}{\hat\rho_2}\ket{\lambda}
\bra{k} \odot\ket{l},
\end{equation}

\noindent where the product basis $\ket{k}\ket{\kappa}$ is used and
indexes $k$, $\kappa$ stand for the first and second quantum
systems, respectively. The operator coefficients $\hat s_{kl}$ in Eq.\
(\ref{genS}) now take the form:
\begin{equation}
\label{sklnew}
\hat s_{kl}=\sum\limits_{\kappa\lambda}\sum\limits_n \bra{n}
\hat s_{k\kappa,l\lambda}
\ket{n}\bra{\kappa}{\hat\rho_2}\ket{\lambda}.
\end{equation}

\noindent Superoperator ${\cal S}$ depends on both the dynamical transformation
${\cal S}_{1+2}$ and the initial state $\hat\rho_2$, and couples the
initial state of the first system with the final state of the second
system.

\subsubsection{One-time coherent information}
\label{subsubsec:one-time}

One-time information quantities can be easily calculated if the
corresponding joint density matrix is known. In the case of a
single system, the corresponding channel is described by the
identity superoperator ${\cal I}$. For the joint input-output
density matrix (\ref{inout}), we get a pure state $\hat \rho_\alpha
= \sum_i \ket{\rho_i}\ket{\rho_i} \sum_j\bra{\rho_j} \bra{\rho_j}$
and then calculate the entropy exchange $S_e=0$ and the coherent
information $I_c= S_{\rm out}=S_{\rm in}$. In the case of two
systems, the input-output density matrix is the joint density
matrix $\hat \rho_{1+2}$, and the corresponding coherent
information in system 2 on system 1 at time $t$ is $I_c(t)
= S[\hat\rho_2(t)]- S[\hat \rho_{1+2}(t)]$. In the case of unitary
dynamics and a pure initial state of the second system, all initial
eigenstates $\ket{i}$ of the first system
transform into the corresponding orthogonal set $\Psi_i(t)$ of the
(1+2) system, so that the joint entropy is time-independent and
the coherent information yields $I_c(t)= S[\hat\rho_2(t)] - S[\hat
\rho_1(0)]$. If the initial state of the first system is also a
pure state, we get simply $I_c(t)=S[\hat \rho_2(t)]$. For the
TLA case, this simply yields $I_c=1$ qubit, if a maximally entangled
state of two-atom qubits is achieved.

\section{TLA in a resonant laser field}
\label{sec:onequbit}

In this section, we discuss the coherent information transfer
between the quantum states of a TLA in a resonant laser field at
two time instants (Fig.\ \ref{fig1}a). Such quantum channel with
pure dephasing in the absence of an external field was considered
in \cite{lloyd96}. In a more general case, coherent information,
based on the joint input-output density matrix (\ref{inout}), can
be readily calculated by using the matrix representation technique
for the relaxation dynamics superoperator. An interesting question
is how the coherent information depends on the applied resonance
field.

The field changes the relaxation rates of the TLA. These rates are
presented with the real parts of the eigenvalues $\lambda_k$ of the
dynamical Liouvillian ${\cal L}={\cal L}_r + {\cal L}_E$ of the TLA,
where ${\cal L}_r$ and ${\cal L}_E$ stand for the relaxation and field
interaction Liouvillians. For simplicity, we will consider here
relaxation caused only by pure dephasing, combined with the laser field
interaction. The corresponding Liouvillian matrix in the basis of $\hat
e_k=\{\hat I, \hat \sigma_3, \hat\sigma_1, \hat\sigma_2\}$ reads
\begin{equation}
\label{Lm}
L=\left(
\begin{array}{cccc}
  0 & 0       & 0       &      0 \\
  0 & 0       & 0       & \Omega \\
  0 & 0       & -\Gamma &      0 \\
  0 & -\Omega & 0       &  -\Gamma
\end{array}
\right),
\end{equation}

\noindent where $\Gamma$ is the pure dephasing rate, $\Omega$ is the
Rabi frequency, and $\hat \sigma_1$, $\hat \sigma_2$, $\hat \sigma_3$
are the Pauli matrices. The eigenvalues of the matrix (\ref{Lm}) can be
readily calculated and are given by
$$
\lambda_k=\{0, -\Gamma,-(\Gamma+ \sqrt{\Gamma^2 -4\Omega^2})/2,
-(\Gamma-\sqrt{\Gamma^2- 4\Omega^2}) /2\}.
$$

\noindent These values are affected by the resonant laser field with
respect to the unperturbed values $0,$ $\Gamma$, which also affects the
resonant fluorescence spectrum of the TLA. At $\Omega>\Gamma/2$ it results
in so-called Mollow-triplet structure, centered at the
transition frequency, which has been predicted theoretically \cite{mollow}
and subsequently confirmed experimentally \cite{ezekil}.

From the information point of view, the resonant laser field might
reduce the coherent information decay rate and, therefore, lead to the
increase of information, although this information gain could intuitively be
expected only from the laser-induced reduction of the
relaxation  rates of the relaxation superoperator ${\cal L}_r$ itself
\cite{pestov73,lisitsa75,burnett82,gJETP83}.

\begin{figure}
\begin{center}
\epsfysize=5.cm \epsfclipon \leavevmode \epsffile{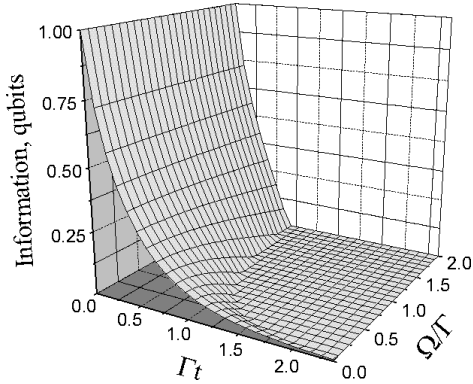}
\end{center}
\caption{The coherent information transmitted between the states of the
TLA at two time instants, $t=0$ and $t>0$, versus time $\Gamma t$ and
the Rabi frequency $\Omega/\Gamma$ (both are dimensionless).}
\label{fig2}
\end{figure}

Calculating the matrix of the evolution superoperator ${\cal
S}=\exp({\cal L}t)$ and using its corresponding representation
(\ref{genS}), the joint density matrix may be calculated
analytically (\ref{inout}). Then (with the help of Eqs (\ref{Se}),
(\ref{Ic})), the coherent information left in the TLA's state at
time $t$ may be calculated about its initial state. This state is
chosen in the form of the maximum entropy density matrix
$\hat\rho_0=\hat I/2$. The results of our calculations are
presented in Fig.\ \ref{fig2}. They show the typical
threshold-type dependence of the coherent information versus time,
which is determined by the loss of coherence in the system. Also,
the coherent information does not increase with an increase of the
laser field intensity, as might be expected. The coherent
information even decreases as the Rabi frequency increases.

In addition, the results show a singularity in the first
derivative of the coherent information dependence at time $t=0$,
which is a characteristic feature of the starting point of the
decay of coherent quantum information. Initially, the input-output
density matrix (\ref{inout}) of the TLA is a pure state
$\hat\rho_\alpha=\Psi\Psi^+$ with the input-output wave function
$\Psi=\sum\sqrt{p_i} \ket{i} \ket{i}$. Its eigenvalues $\lambda_k$
and the probabilities of the corresponding eigenstates are all
equal to zero, except for the eigenstate corresponding to $\Psi$.
Due to the singularity of the entropy function
$-\sum\lambda_k\log\lambda_k$ at $\lambda_k=0$ the derivative of
the corresponding exchange entropy also shows a logarithmic
singularity.

Another interesting feature of coherent information is its
dependence on the initial (input) state $\hat\rho_{in}$. If it were
possible, $\hat\rho_{in}$ might be chosen in the form of the
eigenoperator
$$
\hat\rho_{\rm in}=\sum\limits_{l=1}^4\ket{k_{\min}}_l\hat e_l
$$

\noindent  of the Liouvillian, where $\ket{k_{\min}}$ is the
eigenvector corresponding to the minimum value
$|\Re{e}\lambda_k|>0$ of the matrix $L$. Yet the vector
$\ket{k_{\min}}$ is equal to $\{0, (\Gamma+\sqrt{\Gamma^2 -
4\Omega^2})/2\Omega, 0, 1\}$, which corresponds to the linear
space of operators with zero trace due to the zero value of the
first component. Therefore, the coherent information decay rate
cannot be reduced by reducing the corresponding decay of atomic
coherence.

\section{Coherent information transfer between two subsystems of
the same quantum system}
\label{sec:qubit-intra}

In this section we investigate the quantum channel ($1\to1$, Fig.\
\ref{fig1}a) between two open subsystems $A$ and $B$ of a closed
system $A+B$ having a common Hilbert space ${\rm sp}\,(H_A,H_B)$,
where $H_A$ and $H_B$ are the Hilbert subspaces of the subsystems
$A$ and $B$, respectively.

In classical information theory, this situation corresponds to the
transmission of part $A\subset X$ of the values of an input random
variable $x\in X$. The situation where a receiver receives no message
is also informative and means that $x$ belongs to the supplement
of $A$, $x\in\bar A$. It can be described by the {\em choice}
transformation ${\cal C}=P_A+P_0(1-P_A)$, where $P_A$ is the
projection operator from $X$ onto the subset $A$, $P_A x=x$ for
$x\in A$ and $P_A x=\O$ (empty set) for $x\in\bar A$, $P_0$ is the
projection from $X$ onto an independent single-point set $X_0$,
and $P_0 x=X_0$. This transformation corresponds to the classical
{\em reduction} channel, resulting in information loss only if
$\bar A$ is not a single point. If $\bar A$ is a single point, we are
able to get a maximum of one bit of information, for $\bar A$ can provide
another point of the bit, so that for an input bit we have no loss
of information.

In quantum mechanics, the corresponding reduction channel is
represented as the choice superoperator
\begin{equation}
\label{choice}
{\cal C}=\hat P_A \odot\hat P_A+\ket{0}\bra{0}{\rm Tr}(1-\hat P_A)\odot
(1-\hat P_A),
\end{equation}

\noindent where state $\ket0$ is a quantum analogue of the
classical single-point set, which is separate from all other states.
Eq.\ (\ref{choice}) defines a positive and trace-preserving transformation,
which can appropriately describe coherent information transfer
between subsets of the entire system. The last term in Eq.\
(\ref{choice}) represents the total norm preservation, if all the
states outside the output $B$-set are included. In our case, these
states are included in the incoherent $\ket0\bra0$ form, which in
contrast to the classical one-bit analogue of a TLA yields no coherent
information due to the complete destruction of the coherence.

Considering the coherent information transmitted from part $A$ to
part $B$ of the system, which evolves in time, we deal with the
channel superoperator
\begin{equation}
\label{ABflow}
{\cal S}_{AB}={\cal C}_B{\cal S}_0(t){\cal C}_A,\quad {\cal S}_0(t)=
U(t)\odot U^{-1}(t)
\end{equation}

\noindent with $U(t)$ being the time evolution unitary operator. Here
the input choice superoperator ${\cal C}_A$ is shown just to define the
total channel superoperator, regardless of the input density matrix.
Otherwise, ${\cal C}_A$ is already accounted in the input density matrix
$\hat\rho_{\rm in}$, defined as the operator in the corresponding subspace
$H_A$ of the total Hilbert space $H$.

Let us assume that the dynamical evolution of the system is determined and
the Bohr frequencies $\omega_k$ and the corresponding eigenstates
$\ket{k}$ are found. Then, representing the projectors in terms of the
corresponding input $\ket{\psi_l}$ and output $\ket{\varphi_m}$ wave
functions, Eq.\ (\ref{ABflow}) gives the specified time evolution form
\begin{eqnarray}
\label{SABt}
&{\cal S}_{AB}(t)=\displaystyle \sum_{ll'\in A}\left[\hat
s_{ll'}(t)+\ket0\bra0\sum_{m\notin B}\braket{\varphi_m}{\psi_l(t)}
\braket{\psi_{l'}(t)}{\varphi_m} \right]\bra{\psi_l} \odot\ket{\psi_{l'}},&\nonumber\\
&\hat s_{ll'}(t)= \displaystyle\sum_{mm'\in B} \braket{\varphi_m}{\psi_l(t)}
\braket{\psi_{l'}(t)}{\varphi_{m'}}\,\ket{\varphi_m}\bra{\varphi_{m'}},&\\
&\ket{\psi_l(t)}=\sum_{k}e^{-i\omega_kt} \braket{k}{\psi_l}\ket{k}.&\nonumber
\end{eqnarray}

\noindent Let us consider the case of the orthogonal subsets of
input/output wave functions, which is of special interest. Then, if
there is only one common state $\ket{\phi}$ in the sets $\ket{\psi_l}$,
$\ket{\varphi_m}$ and $U(t_0)=1$ holds for some $t_0$, we get
$$
{\cal
S}_{AB}(t_0)=\ket{\phi} \bra{\phi}\odot\ket{\phi} \bra{\phi}+\ket{0}
\bra{0}\sum \limits_{\varphi_m\ne\phi}\bra{\varphi_m} \odot
\ket{\varphi_m},
$$

\noindent which means that the quantum system is reduced into a
classical bit of the states $\ket{\phi}$ and $\ket0$ and no
coherent information is stored in the subsystem $B$. Nevertheless,
if the eigenstates $\ket{k}$ of $U(t)$ do not coincide with the
input/output states $\ket{\psi_l}$, $\ket{\varphi_m}$ the coherent
information will increase with the time evolution. Hence, the information
capacity of the channel is determined by quantum coupling of the input and
output.

\begin{figure}
\begin{center}
\epsfxsize=5.cm \epsfclipon \leavevmode \epsffile{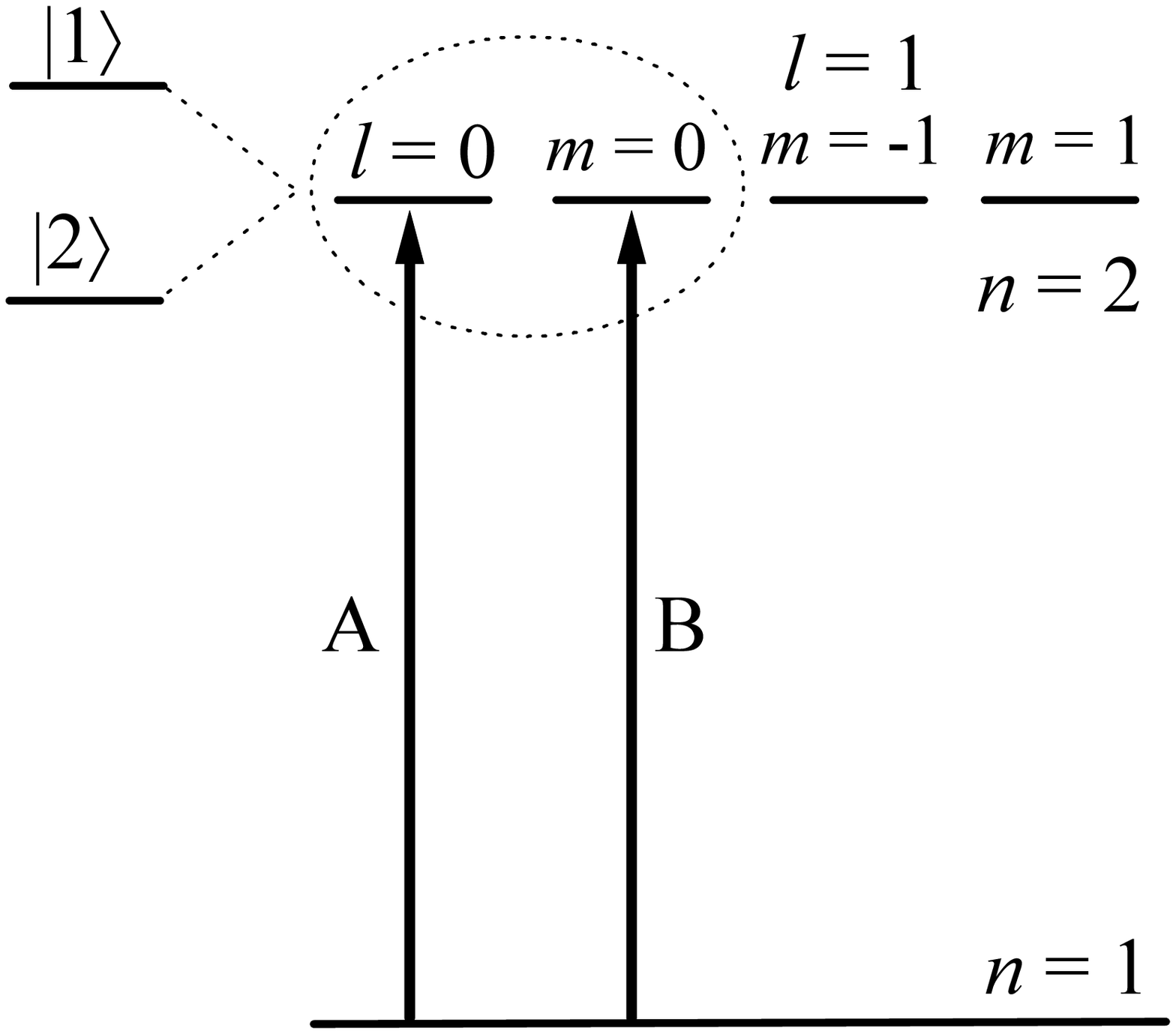}
\end{center}
\caption{A spinless model of the hydrogen atom. The information
channel is made of the input forbidden $nlm\to n'l'm'$ transition
100--200 and the output dipole active 100--210 transition.}
\label{fig3}
\end{figure}

To illustrate the coherent information transfer through the
quantum channel considered in this section, let us analyze a
typical intra-atomic channel between two two-level systems formed
of two pairs of orthogonal states $A=\{\ket{\psi_0},
\ket{\psi_1}\}$ and $B= \{\ket{\psi_0}, \ket{\psi_2}\}$ of the
same atom. A spinless model of the hydrogen atom could serve as
such a system (Fig.\ \ref{fig3}): $\psi_0$ is the ground $s$-state
with $n=1$, $\psi_{1,2}$ are the $s$-state with $l=0,$ $m=0$ and
$p$-state with $l=1,$ $m=0$ of the first excited state with $n=2$,
respectively.

In the absence of an external field, this quantum channel transmits no
coherent information, as the $l=0,$ $m=0$ and $l=1,$ $m=0$ states
are uncoupled. In the presence of an external electric field applied
along the $Z$-axis, the considered two out of four initially
degenerated states with $n=2$ are split, due to the Stark shift
into the new eigenstates $\ket1=(\ket{\psi_1}+ \ket{\psi_2})/
\sqrt2$, $\ket2=(\ket{ \psi_1}- \ket{\psi_2})/\sqrt2$. The
input $l=0$ state oscillates with the Stark shift frequency:
$\ket{\psi_1(t)}= \cos(\omega_st) \ket{\psi_1}+
\sin(\omega_st)\ket{\psi_2}$. Therefore, due to the applied
electric field, the input state becomes coupled to the output state,
which carries the coherent information.

For our model, Eq.\ (\ref{SABt}) presents the $\hat s_{kl}$-operators
in the form of a $3\times3$-matrix, where the third column and row
introduce the phantom ``vacuum'' state $\ket0$:
\begin{eqnarray*}
&\hat s_{11}=\left(\begin{array}{ccc}
  1 & 0 & 0 \\
  0 & 0 & 0 \\
  0 & 0 & 0
\end{array}\right),\quad
\hat s_{12}=\left(\begin{array}{ccc}
  0 & \sin\omega_s t & 0 \\
  0 & 0 & 0 \\
  0 & 0 & 0
\end{array}\right), & \\
&\hat s_{21}=\left(\begin{array}{ccc}
  0 & 0 & 0 \\
  \sin\omega_s t & 0 & 0 \\
  0 & 0 & 0
\end{array}\right),\quad
\hat s_{22}=\left(\begin{array}{ccc}
  0 & 0 & 0 \\
  0 &\sin^2\omega_s t & 0 \\
  0 & 0 &  \cos^2\omega_s t
\end{array}\right).&
\end{eqnarray*}

\noindent Zero values of $\hat s_{12}$, $\hat s_{21}$ correspond to the
absence of coherent information at $t=0$ or to the absence of
coupling. Choosing the input matrix in the maximum entropy form $\hat
\rho_{\rm in}=\hat I/2$, we get the corresponding joint input-output
matrix in the form
$$
\hat\rho_\alpha=\left(\begin{array}{cccccc}
\displaystyle
  \frac{1}{2} & 0 & 0 &\displaystyle \frac{x}{2} & 0 & 0 \\
  0 & 0 & 0 & 0 & 0 & 0 \\
  0 & 0 & 0 & 0 & 0 & 0 \\
\displaystyle  \frac{x}{2} & 0 & 0 & \displaystyle\frac{x^2}{2} & 0 & 0 \\
  0 & 0 & 0 & 0 & 0 & 0 \\
  0 & 0 & 0 & 0 & 0 &\displaystyle \frac{1-x^2}{2}
\end{array}\right),
$$

\noindent where $x=\sin\omega_s t$ and the output density matrix
$\hat\rho_{\rm out}$ is diagonal with the diagonal elements $1/2$,
$x^2/2$, and $(1-x^2)/2$.

Calculating non-zero eigenvalues $(1\pm x^2)/2$ of $\hat\rho_\alpha$
and the entropies $S_{\rm out}$, $S_\alpha$, we get the coherent
information
$$\
I_c=[(1 + x^2)\log_2(1+x^2)-x^2 \log_2(x^2)] /2.
$$

\noindent This function is positive except for $x=0$, where the
coherent information is equal to zero, and its maximum is equal to
1 qubit at $x=\pm1$, e.g., for the precession angle $\omega_s
t=\pm\pi/2$. Thus coherent information on the state of the
forbidden transition is available, in principle, from a dipole
transition via Stark coupling. Its time-averaged value is
$\avr{I_c}=0.46$ qubit.

This forbidden transition was discussed in
\cite{alekseev,moskalev} as a potential source of information on
spatial symmetry breaking caused by the weak neutral current
\cite{weinberg,salam}. For example, if $I_c= 0$, only the
incoherent impact of the forbidden transition (by means of the
ground state population $n_0$) remains and provides a
classical-type of information on the interactions that cannot be
observed directly. In this case, only one
parameter---population---can be potentially measured, while exact
knowledge of the phase of the transition demands $I_c=1$.

\section{Coherent information transfer between two quantum systems}
\label{sec:twoqs}

In recent years, a few results have been published related to
coherent information transfer in a system of two TLAs, including
discussion of the problem from the entanglement measure viewpoint
\cite{hill97} and the ``eavesdropping problem'' \cite{niu99}. A
number of different experiments have been proposed to study
controlled entanglement between two atoms \cite{brennen,trieste}.
From the informational point of view, the coherent information
transmitted in the system of two TLAs connected by a quantum
channel depends both on the specific quantum channel
transformation and the initial states of the TLAs. For the latter,
it seems reasonable to assume that they can be represented by the
product of the independent states of each TLA: $\hat\rho_{1+2}
=\hat\rho_{\rm in} \otimes\hat \rho_{2}$.

In this section, we present a systematic treatment of the coherent
information transfer between two different quantum systems. The
analysis includes coherent information transfer between (i) two unitary
coupled TLAs (subsection \ref{subsec:twoqb}), (ii) two TLAs coupled via
the measuring procedure (subsection \ref{subsec:twoqbm}), (iii) an
arbitrary system and its duplicate (subsection
\ref{subsec:duplication}), (iv) a TLA and the free space photon field
(subsection \ref{subsec:at-f}), and (v) two TLAs coupled via the free space
photon field (subsection \ref{subsec:at-f-at}).

\subsection{Two unitary coupled TLAs}
\label{subsec:twoqb}

Let us first examine a deterministic noiseless quantum channel
connecting two TLAs (Fig.\ \ref{fig1}b). Such a channel can be
described by the unitary two-TLA transformation, which is defined
by the matrix elements $U_{ki,k'i'}$ with $k,i,k',i'=1,2$. Then, the
channel transformation superoperator ${\cal S}$ describing the
transformation $\hat\rho_{\rm in}\to \hat\rho_{\rm
out}=\hat\rho_2'$ can be written in terms of the substitution
symbol (see Eq.\ (\ref{genS})), with operators $\hat{s}_{kl}=
\sum_{\mu\nu}S_{kl,\mu\nu} \ket{\mu}\bra{\nu}$, represented with
the matrix elements of ${\cal S}$ (in accordance with Eqs
(\ref{skl}), (\ref{sklnew})), in the following form:
\begin{equation}
\label{Sklij}
S_{kl,\mu\nu}=\sum_{m\alpha\beta}\rho_{2\alpha\beta}^{}U_{m\mu,k\alpha}^{}
U_{m\nu,l\beta}^*.
\end{equation}

\noindent The relation ${\rm Tr}\,\hat{s}_{kl}=\sum_\mu
S_{kl,\mu\mu} = \delta_{kl}$ is valid here and ensures the correct
normalization condition, whereas the positivity of the block
matrix
$$ (\hat{s}_{kl})=\left(\begin{array}{cc}
  \hat{s}_{11} & \hat{s}_{12}\\
  \hat{s}_{21} & \hat{s}_{22}
\end{array}\right)
$$

\noindent ensures the positivity of ${\cal S}$.

For the no-entanglement transformation $U=U_1\otimes U_2$, Eqs
(\ref{genS}), (\ref{Sklij}) yield ${\cal S}=\hat\rho_{2}'{\rm
Tr}\;\odot$, which means that the initial state $\hat\rho_1$ of
the first TLA transfers into the final state, which is not
entangled with the state $\hat\rho_2'=U_2 \hat\rho_2U_2^+$ of the
second TLA.

We can simplify Eq.\ (\ref{Sklij}) by considering a pure state
$\hat\rho_2$, so that together with an arbitrary choice of
no-entanglement transformation $U$, it seems reasonable to consider
a special case of the pure state: $\rho_{2\alpha\beta}=
\delta_{\alpha\beta} \delta_{\alpha\alpha_0}$. Keeping also in
mind that $S_{kl,\mu\nu}$ is linear on the density matrix $\hat\rho_2$
and the coherent information $I_c$ is a convex function of $\cal
S$ \cite{barnum98}, Eq.\ (\ref{Sklij}) simplifies to
\begin{equation}
\label{Sklij1}
S_{kl,\mu\nu}=\sum_{m}U_{m\mu,k\alpha_0}^{} U_{m\nu,l\alpha_0}^*,
\end{equation}

\noindent which means that the quantum channel is described only by
the unitary transformation $U$. Here the summation is taken over only
the states $\ket{m}$ of the first TLA after the coupling
transformation.

\begin{figure}
\begin{center}
\epsfxsize=6.cm\epsfclipon\leavevmode\epsffile{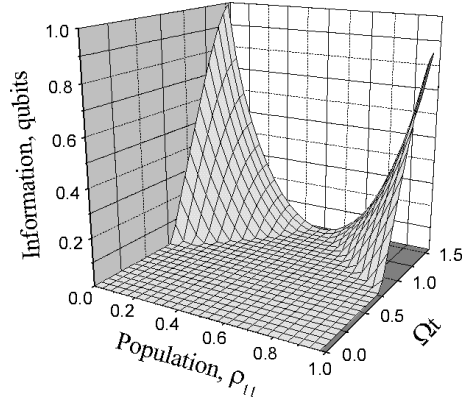}
\end{center}
\caption{The coherent information transmitted between two unitary
coupled TLAs versus population $\rho_{11}$ of the diagonal initial
density matrix of the second TLA and the coupling precession angle
$\Omega t$.} \label{fig4}
\end{figure}

The coherent information transmitted in systems of two unitary
coupled TLAs with $\hat\rho_{\rm in}=\hat I/2$ and
$\left(\hat\rho_2\right)_{12}=\sqrt{\left(\hat\rho_2\right)_{11}\left[1-
\left(\hat\rho_2\right)_{11}\right]}$ is shown in Fig.\
\ref{fig4}. A convex function of $\hat\rho_2$ is shown, which has
the maximum on the border,
$\rho_{11}=\left(\hat\rho_2\right)_{11}=0,1$. As in the case of a
single TLA, the behavior of the coherent information preserves the
typical threshold-type dependency on the coupling angle, which
determines the degree of the coherent coupling of two TLAs with
respect to the independent fluctuations of the second TLA.

\subsection{Two TLAs coupled via the measuring procedure}
\label{subsec:twoqbm}

Here we will discuss a specific type of quantum channel
connecting two TLAs \cite{TTLA}, where the
superoperator $\cal S$ is defined by the {\em measuring procedure}, which
implements a different approach to the quantum information
\cite{chen99} called {\em measured information}.

We start with a channel formed of two identical two-level
systems. In terms of wave function, the corresponding {\em full
measurement} transformation of the first TLA state is defined as
\begin{equation}
\label{meas}
\psi\otimes\varphi\to\sum a_i\ket{\phi_i}\ket{\phi_i},\quad
a_i=\braket{\phi_i}{\psi}.
\end{equation}

\noindent This transformation provides full entanglement of some basis states
$\ket{\phi_i}$, which do not depend on the initial state $\varphi$
of the second TLA. The latter serves as a measuring device, yet
fully preserves information on the basis states of the first
system state $\psi=\sum a_i\ket{\phi_i}$. Eq.\ (\ref{meas}), being
a deterministic transformation of the wave function, is
neither linear nor unitary transformation with respect to $\varphi$ and,
therefore, cannot represent a true deterministic transformation.
The corresponding representation in terms of the two-TLA density
matrices has the form:
\begin{equation}
\label{rho12}
\hat\rho_{12}\to\sum\limits_i\sum\limits_j\bra{\phi_i}\bra{\phi_j}\hat
\rho_{12}\ket{\phi_j}\ket{\phi_i}\, \ket{\phi_i}\ket{\phi_i}\bra{\phi_i}
\bra{\phi_i}.
\end{equation}

\noindent This representation is linear on $\hat\rho_{12}$ and satisfies the
standard conditions of physical feasibility
\cite{barnum98,kraus83}, i.e completely positive and trace
preserving. This matrix is in the form of $\sum p_i\ket{\phi_i}
\ket{\phi_i}\bra{\phi_i}\bra{\phi_i}$, so that $S(\hat
\rho_{12})=S(\hat\rho_2)$. Due to the classical nature of the
information represented here only with the classical indexes $i$
and in accordance with the equations of section \ref{sec:definitions}, the
single-instant coherent information is zero.

In the case of a two-time channel, the superoperator for the quantum channel
connecting two TLAs can be readily derived from Eq.\ (\ref{genS})
with $\hat s_{kl}=\ket{\phi_k}\bra{\phi_k} \delta_{kl}$,
$\bra{k}\to\bra{\phi_k}$, and $\ket{k}\to\ket{\phi_k}$. After
calculating the trace over the first TLA and replacing
$\hat\rho_{12}$ with the substitution symbol $\odot$, the
equation takes the form:
\begin{equation}
\label{dirmeas}
{\cal M}=\sum\limits_k\hat P_k {\rm Tr}_1^{}\hat E_k\,\odot.
\end{equation}

\noindent Here $\hat P_k=\ket{\phi_k}\bra{\phi_k}$ are the orthogonal
projectors representing the eigenstates of the ``pointer'' variable of
the second TLA and $\hat E_k=\ket{\phi_k}\bra{\phi_k}$ is the
orthogonal expansion of the unit (orthogonal map) formed of the same projectors.
This orthogonal map determines here the quantum-to-classical reduction
transformation ${\rm Tr}_1\hat E_k \, \odot= \bra{\phi_k}
\odot \ket{\phi_k}$, which represents the procedure of getting classical
information $k$ from the first system.
Applying the transformation (\ref{dirmeas}) to $\hat\rho_{\rm
in}$ and using Eq.\ (\ref{inout}) for the respective output and
input-output density matrices, we get
\begin{equation}
\label{rhodirmeas}
\hat\rho_{\rm out}= \sum\limits_k \tilde p_k \ket{\phi_k}
\bra{\phi_k},\quad \hat\rho_\alpha= \sum\limits_k \tilde p_k
\ket{\phi_k} \ket{\pi_k} \bra{\pi_k}\bra{\phi_k},
\end{equation}

\noindent where $\tilde p_k= \bra{\phi_k} \hat\rho_{\rm in}\ket{\phi_k}
= \sum_i p_i |\braket{\phi_k} {i}|^2$ are the eigenvalues
of the reduced density matrix and $\ket{\pi_k}= \sum_i\sqrt{p_i/\tilde
p_k}\braket{\phi_k} {i}\ket{\bar{i}}$ are the normalized modified input
states coupled with the output states $\ket{\phi_k}$ after the
measurement procedure. It is important to note (as it follows from
Eq.\ (\ref{rhodirmeas})) that there is no coherent information in the system
because vectors $\ket{\phi_k}$ are orthogonal and therefore the
entropies of the density matrices (\ref{rhodirmeas}) are obviously the
same. Conversely, the measured information introduced in \cite{chen99}
is not equal to zero in this case.

We can easily generalize our result for a more general case of the
quantum channel, when the second system has a different structure
from the first and, therefore, they occupy different Hilbert
spaces. This difference leads to the replacement of the basis states
$\ket{\phi_i}$ of the second system in our previous results with
another orthogonal set $\ket{\varphi_i}= V\ket{\phi_i}$, where $V$
is an isometric transformation from the Hilbert state $H_1$ of the
first system to the different Hilbert space $H_2$ of the second
system. After simple straightforward calculations, the final
result is the same---there is no coherent information
transmitted through the quantum channel. This result is a natural
feature of coherent information, in contrast to other information
approaches (see, for example Ref. \cite{chen99}).

It is interesting to discuss more general measuring-type
transformations, for instance, the indirect (generalized)
measurement procedure. This procedure was first applied to the
problems of optimal quantum detection and measurement in
\cite{helstr70} and then, in a form of non-orthogonal
expansion of unit $\hat{\cal E} (d\lambda)$, in \cite{gTK73}
($\hat{\cal E} (d\lambda)$ is equivalent to the positive
operator-valued measure, POVM, used in the semiclassical version of
quantum information and measurement theory
\cite{preskill,helstrom,peres93}). This indirect measuring
transformation results from averaging a direct measuring
transformation applied, not to the system of interest, but to its
combination with an auxiliary independent system. The
indirect-measurement superoperator in the general form can be written
as
\begin{equation}
\label{indmf}
{\cal M}=\sum\limits_q\hat P_q{\rm Tr}\,\hat{\cal E}_q\odot,
\end{equation}

\noindent where $\hat P_q$ are the arbitrary orthogonal projectors and
$\hat{\cal E}_q$ is the general-type non-orthogonal expansion of the
unit in $H$ space (POVM). Note that $\hat{\cal E}_q=\ket{\varphi_q}
\bra{\varphi_q}$ is a specific ``pure'' type of POVM, first used in
quantum detection and estimation theory \cite{helstr70}. The latter
describes the full measurement in $H\otimes H_a$ for the
singular choice of the initial auxiliary system density matrix
$\rho^a_{bc}= \delta_{b0} \delta_{bc}$.

The information transfer from the initial density matrix to the final
output state is represented in Eq.\ (\ref{indmf}) via the coupling
provided by indexes $q$. Because the number $N_q$ of $q$ values can be
greater than ${\rm Dim}\,H$, it seems reasonable to suggest that some
output coherent information is left about the input state. The
corresponding output and input-output density matrices are given by
\begin{equation}
\label{rhoalpha}
\hat\rho_{\rm out}=\sum\limits_q
\tilde p_q \hat P_q,\quad \hat\rho_\alpha=\sum\limits_{qij} \sqrt{p_i
p_j}\bra{j}\hat{\cal E}_q\ket{i}\hat P_q\otimes\ket{\bar{i}}\bra{\bar{j}},
\end{equation}

\noindent where $\tilde p_q={\rm Tr}\,\hat{\cal E}_q\hat\rho_{\rm in}$
are the state probabilities given by the indirect measurement.

In the case of full indirect measurement, it can be easily
inferred theoretically or confirmed by numerical calculations for
particular examples that no coherent information is available. The
proof is based on the quantum analogue \cite{schum96} of the
classical data processing theorem and the above discussed result
on a full direct measurement. Therefore, in order to get non-zero
coherent information, a class of incomplete (soft) measurements
must be implemented, which are subject to more detailed quantum
information analyzis.

\subsection{Quantum duplication procedure}
\label{subsec:duplication}

In the previous subsection, we demonstrated that the classical-type
measuring procedure defined by the transformation (\ref{rho12})
completely destroys the coherent information transmitted through the
quantum channel. Here we will consider a modified transformation for
the quantum channel shown in Fig.\ \ref{fig1}c, which preserves the coherent
information:
$$
\hat\rho_{12} \to\hat\rho_{12}'= \sum \limits_{ij} \bra{\phi_i} {\rm
Tr}_2\hat\rho_{12}\ket{\phi_j} \ket{\phi_i}\ket{\phi_i}
\bra{\phi_j}\bra{\phi_j}.
$$

\noindent In this equation off-diagonal matrix elements of the input density
matrix $\hat\rho_1=\hat\rho_{\rm in}$ are taken into account, which
preserves the phase connections between different $\phi_i$.

For the initial density matrix of a product type $\hat\rho_{\rm
in}\otimes \hat\rho_2$, in terms of $\hat \rho_{\rm in} \to\hat
\rho_{12}'$ transformation from $H$ to $H\otimes H$, the corresponding
superoperator has the form:
\begin{equation}
\label{rho12coh}
{\cal Q}=\sum\limits_{ij} \ket{\phi_i}\ket{\phi_i}\bra{\phi_j} \bra{\phi_j}
\bra{\phi_i}\odot\ket{\phi_j}.
\end{equation}

\noindent This superoperator defines the coherent measuring
transformation, in contrast to the incoherent transformation discussed in
\cite{chen99}. The coherent measuring transformation converts
$\hat\rho_{\rm in}$ into $\hat\rho_2$-independent state
\begin{equation}\label{rhooutQ}
\hat\rho_{\rm out}=\rho_{12}'=\sum_{ij} \bra{\phi_i}\hat\rho_{\rm in}
\ket{\phi_j} \ket{\phi_i}\ket{\phi_i} \bra{\phi_j} \bra{\phi_j},
\end{equation}

\noindent which results in duplication of the input eigenstates
$\phi_i$ into the same states of the pointer variable $\hat k=\sum_k k
\ket{\phi_k}\bra{\phi_k}$. Pure states of the input are transformed
into the pure states of the joint (1+2)-system by doubling the pointer
states:
$$
\psi\to\sum\limits_i \braket{\phi_i}{\psi}\ket{\phi_i}\ket{\phi_i}.
$$

\noindent This mapping is similar to the mapping given by Eq.\
(\ref{meas}). Of course, only the input states $\psi$ equal to the
chosen pointer basis states $\phi_k$ are duplicated without
distortion because it is impossible to transmit non-orthogonal
states using only orthogonal ones. The entropy of the output state
with a density matrix (\ref{rhooutQ}) having the same matrix
elements as $\hat\rho_{\rm in}$, is evidently the same as the
input state, $S_{\rm out}=S_{\rm in}=S[\hat\rho_{\rm in}]$, due to
the preservation of the coherence of all pure input states.

For the joint input-output states, the transformation (\ref{rho12coh})
yields the corresponding density matrix (\ref{inout}) in $H\otimes
H\otimes H$ space:
\begin{equation}
\label{rhoalphaQ}
\hat\rho_\alpha=\sum\limits_{kl}\ket{\phi_k}\ket{\phi_k}\bra{\phi_l}
\bra{\phi_l} \otimes\sqrt{\tilde p_k\tilde p_l}\ket{\chi_k}\bra{\chi_l},
\end{equation}

\noindent where $\tilde p_k$, $\ket{\chi_k}$ are the same as
above, providing an expansion of the input density matrix in the
form $\hat\rho_{\rm in}=\sum_k \tilde p_k \ket{\chi_k}
\bra{\chi_k}$. Taking into account that the first tensor product
term in Eq.\ (\ref{rhoalphaQ}) is a set of transition projectors
$\hat P_{kl},\, \hat P_{kl}\hat P_{mn}=\delta_{lm}\hat P_{kn}$, we
can apply easily proven algebraic rules valid for a scalar
function $f$: $$ f(\sum_{kl}\hat P_{kl} \otimes \hat
R_{kl})=\sum_{kl}\hat P_{kl}\otimes f(\hat R)_{kl}, $$

\noindent where $\hat R=(\hat R_{kl})$ is the block matrix and
${\rm Tr}\, f(\sum_{kl}\hat P_{kl} \otimes \hat R_{kl})={\rm
Tr}\,f(\hat R)$. Here $\hat R= \left(\sqrt{\tilde p_k\tilde
p_l}\ket{\chi_k} \bra{\chi_l}\right)$, and it is simply
$\ket{\ket{\chi}}\bra{\bra{\chi}}^+$ with $\ket{\ket{\chi}}_{ki} =
\sqrt{\tilde p_k}\chi_{ki}$, a vector in the $H\otimes H$ space.
All eigenvalues $\lambda_k$ of this matrix are equal to zero,
except one value corresponding to the eigenvector
$\ket{\ket{\chi}}$.

Calculation of the exchange entropy gives $S_e=0$, and, therefore,
$I_c=S_{\rm in}$. Consequently, the coherent duplication does not
reduce the input information transmitted through the 1$\to$(1+2)
channel, nor does it matter whether the register $\hat k$ is
compatible with the input density matrix, $[\hat k,\hat\rho_{\rm
in}]=0$, or not.

If the channel is reduced to the one shown in Fig.\ \ref{fig1}b
and discussed in the previous subsection, by taking in Eq.\
(\ref{rhooutQ}) trace either over the first or the second system,
we evidently come to the measurement procedure discussed in
subsection \ref{subsec:twoqbm}. As a result, we can conclude that
the coherent information is strictly associated with the joint
system but not with its subsystems. This natural property could
be used in quantum error correction algorithms
\cite{errorcorr} or for producing stable entangled states
\cite{lasphys}.

\subsection{TLA-to-vacuum field channel}
\label{subsec:at-f}

In this subsection, we analyze the quantum channel between a TLA
and a vacuum electromagnetic field (Fig.\ \ref{fig1}b), which is an
extension of the TLA in an external laser field, as considered in
section \ref{sec:onequbit}.

\begin{figure}
\begin{center}
\epsfxsize=5.5cm \epsfclipon\leavevmode\epsffile{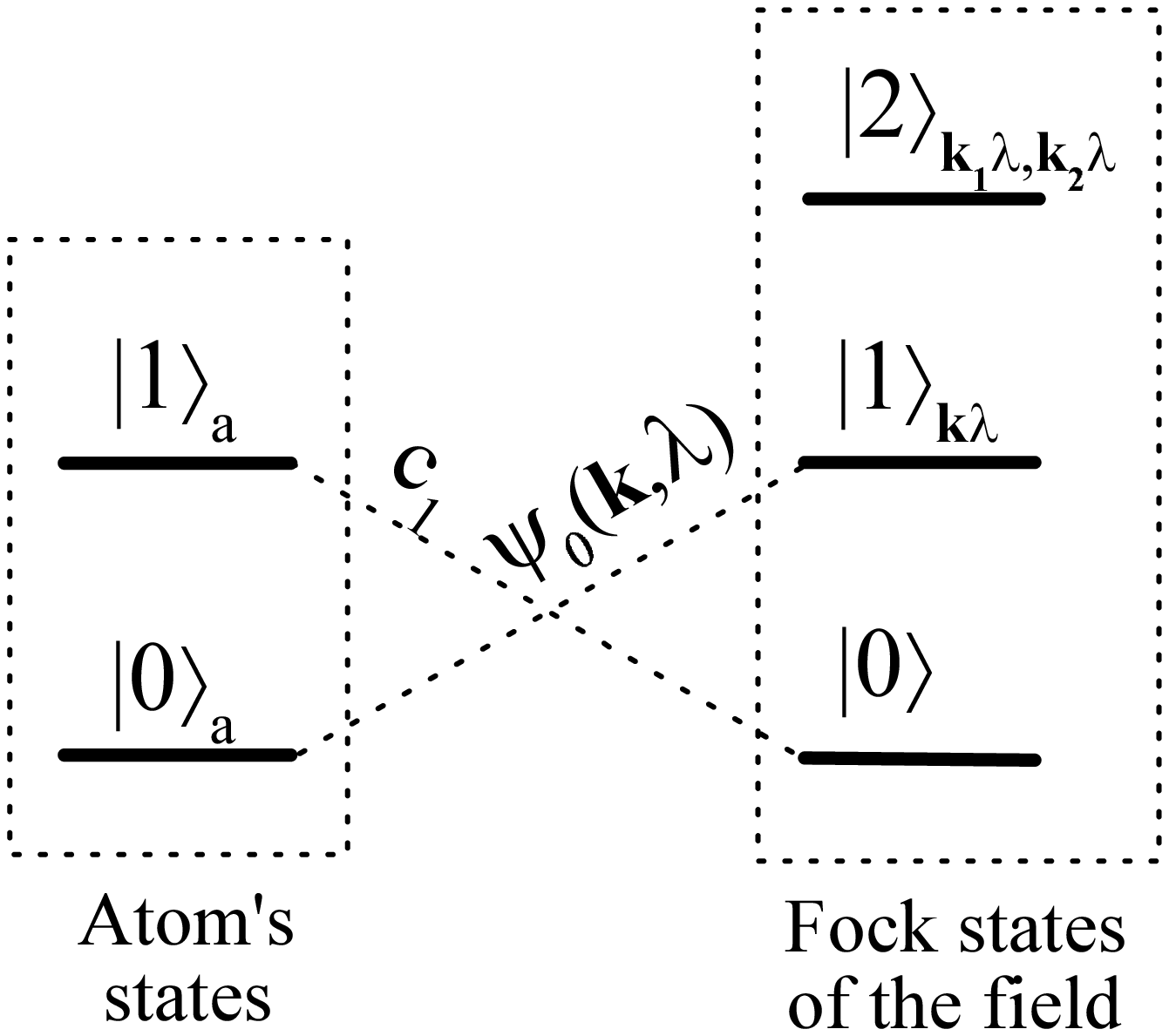}
\end{center}
\caption{Structure of the joint Hilbert space of the (atom+field)
system. For the vacuum initial field state, both atomic states and only
two Fock states of the field ($\ket0$ and $\ket1$) are involved in the
dynamics of the joint system (atom+field). The dynamics is entirely
defined by just two states, $\ket0_a\ket1_{{\bf k}\lambda}$ and
$\ket1_a\ket0$, which are described by $\psi_0({\bf k},\lambda)$ and
$c_1$, respectively.}
\label{fig5}
\end{figure}

For this analysis, we will use a reduced model of the field, which
is based on the reduction of the Hilbert space of the field in the
Fock representation (Fig.\ \ref{fig5}). The problem, therefore, is
reduced to that of the interaction of a two-level system with
continuous multi-mode oscillator systems \cite{Cohen92}, a
specific case of which is the interaction of an atom with the free
photon field. However, to analyze the information in the system
(atom+field), we do not need to consider the specific dependence
of the wave function $\psi_0({\bf k},\lambda)$ of the field photon
on the wave vector (including polarization), because only its
total probability and phase are significant.

In the basis of the free atomic and field states for the vacuum's initial
state $\alpha_0=0$, we get from Eq.\ (\ref{Sklij1})
$$
S_{kl,\mu\nu}=\sum_m U_{m\mu,k0}^{} U_{m\nu,l0}^*.
$$

\noindent Greek letters are used to distinguish the photon field
indexes, which in the general case include both the number of
photons and their space or momentum coordinates. Matrix elements
of this superoperator calculated via the atom-to-field unitary
evolution matrix $U_{m\mu,k0}$ coefficients (Table \ref{table1})
are shown in Table \ref{table2}.

The choice of $\psi_0({\bf k},\lambda)$ as a basis for the photon
field \cite{addnote} reduces the matrix of operator $S_{kl,\mu\nu}$ to the
non-operator matrix transformation, which in terms of $\hat
s_{kl}$ matrices has the form:
\begin{equation}
\label{Sklmnr}
\begin{array}{cc}
\hat s_{11}=
\left(\begin{array}{cc}1&0\\0&0\end{array}\right),&
\hat s_{12}=
\left(\begin{array}{cc}0&\displaystyle\left(1-e^{-\gamma t}\right)^{1/2}\\
0&0\end{array}\right),\\
\hat s_{21}=
\left(\begin{array}{cc}0&0\\\displaystyle\left(1-e^{-\gamma t}
\right)^{1/2}&0\end{array}\right),&
\hat s_{22}=
\left(\begin{array}{cc}e^{-\gamma t}&0\\0&1-e^{-\gamma t}\end{array}\right),
\end{array}
\end{equation}

\noindent where $|c_1|^2=\exp(-\gamma t)$ describes the population
decay of the totally populated initial excited state of the atom and
$\int\sum |\psi_0({\bf k},\lambda)|^2 {\rm d}{\bf k}=1-\exp(-\gamma t)$ is the
probability a photon will be detected. From Eq.\ (\ref{Sklmnr}), it follows
that the structure of the photon field plays no role, and the transmitted
information defined by the input-output density matrix depends only on
the photon emission probability by time $t$. The reduction of the
photon field (only the photon numbers $\mu,\nu=0,1$ were taken into
account) leads to the  conclusion that the photon states also are equivalent to
those of a two-level system.

Applying the transformation (\ref{Sklmnr}) to the input atom density matrix
$$
\hat\rho_{\rm in}=
\left(\begin{array}{cc}
  \rho_{11} & \rho_{12} \\
  \rho_{12} & 1-\rho_{11}
\end{array}\right),
$$

\noindent restricted to the real off-diagonal matrix elements, we get
the output density matrix
$$ \hat\rho_{\rm out}=\left(\begin{array}{cc}
  \rho_{11}+\rho_{22}e^{-\gamma t} & \rho_{12}
  \displaystyle\left(1-e^{-\gamma t}\right)^{1/2} \\
  \rho_{12}\displaystyle\left(1-e^{-\gamma t}\right)^{1/2} &
  \rho_{22}(1-e^{-\gamma t})
\end{array}\right)
$$

\noindent and for $\rho_{12}=0$ the respective input-output density matrix
$$
\hat\rho_\alpha= \left(
\begin{array}{cc|cc}
\rho_{11} & 0 & 0 & \displaystyle\left[\rho_{11}\rho_{22}
(1-e^{-\gamma t})\right]^{1/2}\\
0 & \rho_{22}e^{-\gamma t} & 0 & 0 \\ \hline
0 & 0 & 0 & 0 \\
\displaystyle\left[\rho_{11}\rho_{22}(1-e^{-\gamma t})
\right]^{1/2} & 0 & 0 & \rho_{22}(1-e^{-\gamma t})
\end{array}\right).
$$

For $t\to\infty$ this expression yields a pure atom-photon state,
which converts incoherent fluctuations of the atomic states,
forming the incoherent ensemble, to equivalent coherent
fluctuations of the photon states. The corresponding eigenvalues
are $\lambda_\alpha= \{0,0,1-\rho_{22}\exp(-\gamma t),\rho_{22}
\exp(-\gamma t)\}$. Non-zero values are equal to the probabilities
of the atomic states at time $t$. For the output (photon) density
matrix $\hat\rho_{\rm out}$ the eigenvalues are $\lambda_{\rm
out}=\{\rho_{22}[1-\exp(-\gamma t)],1-\rho_{22}[1- \exp(-\gamma
t)]\}$, which are the probability that a photon will be emitted or
not. These sets of eigenvalues determine the eigen probabilities
of the joint input-output and marginal output matrices. The
coherent information, defined by the difference of the
corresponding entropies, then takes the form:
\begin{equation}
\label{Ican}
\begin{array}{ll}
I_c = & x\rho_{\rm _{22}} \log_2(x \rho_{22})-(1-\rho_{22}+x \rho_{22})
\log_2[1-(1-x) \rho_{22}]+\\ &(1-x \rho_{22}) \log_2(1-x \rho_{22})-
(1-x)\rho_{22} \log_2(\rho_{22}-x \rho_{22}),
\end{array}
\end{equation}

\noindent where $x=\exp(-\gamma t)$. This formula is valid for $I_c>0$,
otherwise, $I_c=0$. The corresponding critical point is $\exp(-\gamma
t)= 1/2$, the time when the probability $1-\rho_{22} [1- \exp(-\gamma
t)]$ of finding no photon is equal to the population of the lower
atomic state $1-\rho_{22}\exp(-\gamma t)$.

\begin{figure}
\begin{center}
\epsfxsize=6.cm\epsfclipon\leavevmode\epsffile{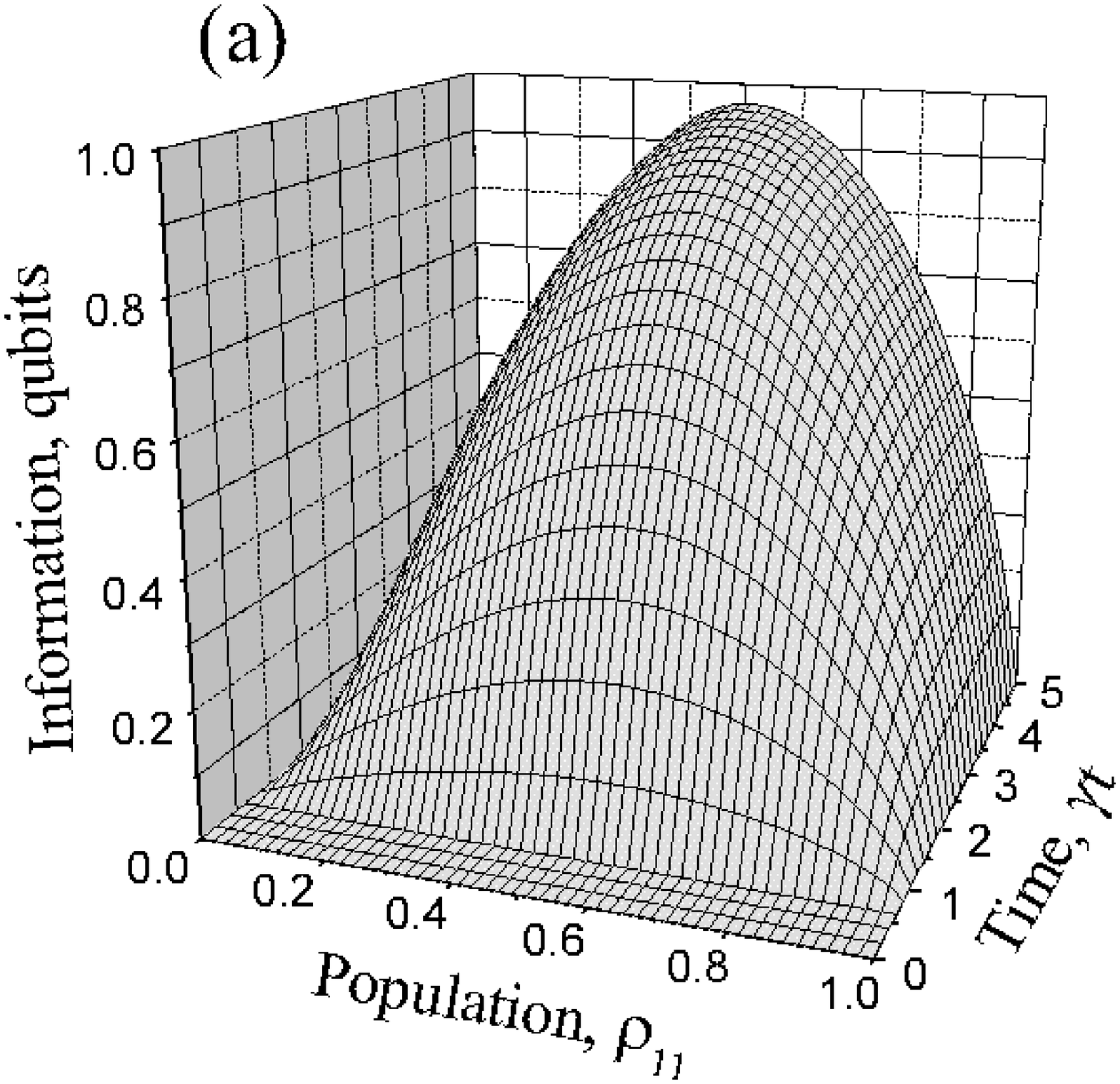}
\epsfxsize=6.cm\epsfclipon\leavevmode\epsffile{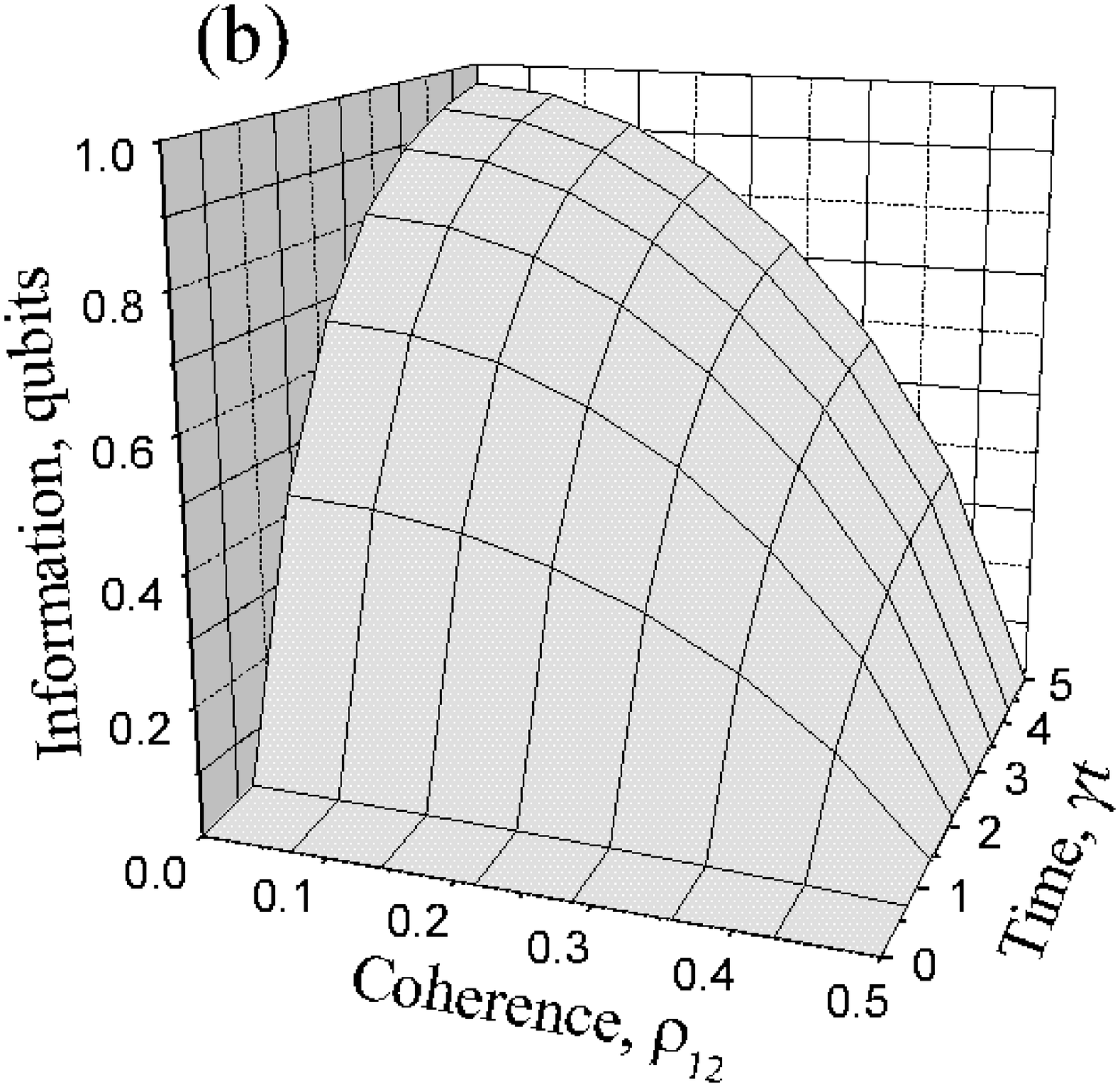}
\end{center}
\caption{The coherent information transmitted in the atom-to-field
quantum channel versus the dimensionless time $\gamma t$ and input
atomic density matrix, which is taken either diagonal with the ground
state matrix element $\rho_{11}$ (a) or as the sum of $\hat I/2$ and the real
(``cosine-type'') coherent contribution of the off-diagonal elements
$\rho_{12}\hat\sigma_1$ (b).}
\label{fig6}
\end{figure}

The results for calculating the coherent information are shown in
Fig.\ \ref{fig6} for two specific cases: $\rho_{12}=0$ (Fig.\
\ref{fig6}a) and $\rho_{11}=1/2$, $0\le\rho_{12}\le1/2$ (Fig.\
\ref{fig6}b). One can see from Fig.\ \ref{fig6}a that the coherent
information is symmetrical with respect to the population
$\rho_{11}$ around the symmetry point $\rho_{11}=1/2$. Increasing
the excited state population $\rho_{22}=1-\rho_{11}$ and the
corresponding photon emission yield does not increase the coherent
information, because of the reduction of the source
entropy, which determines the potential maximum value of the
coherent information. For the same reason, the coherent
information decreases when there is a non-zero coherent
contribution to the initial maximum entropy atom state and
completely vanishes for the pure coherent initial state (Fig.\
\ref{fig6}b).

In accordance with section \ref{sec:definitions} and because of
the purity of the initial field state, one-time information
is equal to the difference of the entropies of the photon field
only, represented by $\hat\rho_{\rm out}$, and the initial atomic
state, represented by $\hat\rho_{\rm in}$. For a pure initial
state, expressed in the form of the excited atom state $\ket2$,
and for $0<t<\infty$, we always get non-zero information
$I_c=-x\log_2 x-(1-x)\log_2 (1-x)$ that yields 1 qubit for
$x=1/2$, when the excited state population is equal to the
probability a photon will be emitted.

\subsection{The transmission of coherent information between two atoms
via a free space field}
\label{subsec:at-f-at}

In this subsection, we will consider the quantum channel when
information is transmitted from one atom to another via the free
space field (Fig.\ \ref{fig1}b). Suppose that the second atom is
initially in the ground state. In addition, we will restrict
ourselves here to the long time scale approximation, in which the
effects of the discrete nature of the retarding electromagnetic
interaction are neglected
\cite{fermi32,hamilton49,heitler49,milonni74}. Under such
restrictions and approximations we have the Dicke problem
\cite{dicke54}, for which the well-known solution for the atomic
state in the form of two decaying symmetric and antisymmetric
Dicke states $\ket{s}=(\ket1\ket2+\ket2\ket1)/\sqrt2,$
$\ket{a}=(\ket1\ket2- \ket2\ket1)/\sqrt2$ and the stable vacuum
state $\ket0=\ket1\ket1$ can be written as:
\begin{eqnarray}
\label{dickedyn} &c_s(t)=c_s(0)\exp[-(\gamma_s/2+i\Lambda)t],&\nonumber\\
&c_a(t)=c_a(0)\exp[-(\gamma_a/2-i\Lambda)t],&\\
&c_0(t)=c_0(0)+\left[c_s(0)^2+c_a(0)^2-c_s(t)^2-c_a(t)^2\right]^{1/2}
e^{i\xi(t)}.&\nonumber
\end{eqnarray}

\noindent Here $c_0(t)$ is the amplitude of the stable vacuum
component $\ket1\ket1$, which has an incoherent contribution due to  the
spontaneous radiation transitions from the excited two-atomic
states, $\xi(t)$ is the homogeneously distributed random phase,
$\gamma_{s,a}$ and $\Lambda$ are their decay rate and coupling
shift, respectively, and $c_{s,a}$ are the amplitudes of the Dicke
states.

In terms of the products of the individual atomic states
$\ket{i}\ket{j}$ for the corresponding initial amplitudes
$c_{12}(0)=0,$ $c_{22}(0)=0$ the system's dynamics is described,
according to the Dicke dynamics (\ref{dickedyn}), by the following
equations:
\begin{eqnarray*}
&c_{11}(t)=c_{11}(0)+f(t)e^{i\xi(t)}c_{21}(0), \quad c_{21}(t)=
f_s(t)c_{21}(0),&\\ & c_{12}(t)=f_a(t)c_{12}(0),\quad c_{22}(t)=0,&\\
&f(t)=\left\{1-[\exp(- \gamma_st)+ \exp(-\gamma_at)]/2\right\}^{1/2},&\\
&f_s(t)= \{\exp[-(\gamma_s/2+ i\Lambda)t]+\exp[-(\gamma_a/2-i\Lambda)t]\}/2,&\\
&f_a(t)=\{\exp[-(\gamma_s/2 +i\Lambda)t]-\exp[-(\gamma_a/2-i\Lambda)t]\}/2.&
\end{eqnarray*}

\noindent Applying these formulas to the input operators
$c_{k1}(0)c_{l1}^*(0) \ket{k}\bra{l}$ of the first atom and then
averaging the output over the final states of the first atom and the
field fluctuations (the latter is represented here only with $\xi(t)$),
we get the symbolic channel superoperator transformation $\hat
\rho^{(1)} (0)\to\hat\rho^{(2)}(t)={\cal S}(t)\hat\rho^{(1)}(0)$ and
corresponding $\hat s_{kl}$ operators in the form:
\begin{eqnarray}
\label{at-f-at-dyn}
&\begin{array}{ccl}
{\cal S}(t)&=&\ket1\bra1\odot\ket1\bra1+\left[f(t)^2+|f_s(t)|^2\right]\ket1
\bra2\odot\ket2\bra1+|f_a(t)|^2\ket2\bra2\odot\ket2\bra2\\
 & &+f_a(t)\ket2\bra2\odot\ket1\bra1+f_a^*(t)\ket1\bra1\odot\ket2\bra2,
\end{array}& \nonumber \\
&\hat s_{11}=\left(\begin{array}{cc}1&0\\0&0\end{array}\right),\quad
\hat s_{12}=\left(\begin{array}{cc}0&f_a^*(t)\\0&0\end{array}\right),  &\\
&\hat s_{21}=\left(\begin{array}{cc}0&0\\f_a(t)&0\end{array}\right),\quad
\hat s_{22}=\left(\begin{array}{cc}f(t)^2+|f_s(t)|^2&0\\0&|f_a(t)|^2 \nonumber
\end{array}\right).
\end{eqnarray}

To further elucidate this problem, let us now discuss the case of two
identical atoms having parallel dipole moments aligned perpendicular
to the vector connecting the atoms. Here only two dimensionless
parameters are essential: dimensionless time, $\gamma t$, where
$\gamma$ is the free atom's decay rate, and dimensionless distance,
$\varphi=k_0R$, where $R$ is the interatomic distance and $k_0$ is the
wave vector at the atomic frequency. Then, the dimensionless two-atomic
decay rates and the short distance dipole-dipole shift are given by
\cite{trieste,lasphys,milonni74}:
$$
\gamma_{s,a}/\gamma=1\pm g\quad \text{and}\quad
\Lambda/\gamma= (3/4)/\varphi^3,
$$

\noindent respectively, with $g=(3/2)(\varphi^{-1} \sin\varphi +
\varphi^{-2} \cos\varphi- \varphi^{-3}\sin\varphi)$.

The coherent information may be calculated as previously described in subsection
\ref{subsec:at-f} by replacing $\exp(-\gamma t)$ with
$f(t)^2+|f_s(t)|^2$ in Eq.\ (\ref{Sklmnr}). Then, the operators $\hat
s_{kl}$ in Eq.\ (\ref{Sklmnr}) become similar to the corresponding
operators in Eq.\ (\ref{at-f-at-dyn}). The coherent information is
given by the same Eq.\ (\ref{Ican}) with $x=f(t)^2+ |f_s(t)|^2$, which,
however, now has (in contrast with a single-atom case considered in
\ref{subsec:at-f}) new qualitative features arising from the specific
oscillatory dependence of $|f_{s,a}(t)|^2$ on the interatomic distance
$\varphi$.

If there were no oscillations from the quasi-electrostatic
dipole-dipole coupling, i.e. as in the case of $\Lambda=0$, the
coherent information would always be equal to zero, because the
threshold $x<0.5$ would not be achieved. Parameter $(1-x)$
corresponds to the population of the excited state of the second
atom for the initial state $\ket2$ of the first atom, and for the
optimal value $\rho_{22}=1/2$ of its initial population (from the
information point of view), we have $1-x\le1/4$ and $x\ge3/4$.
Oscillations in $|f_a(t)|^2$ lead to the interference between the
two decaying Dicke components, so that the maximum of the
population $n_2=1-x$ goes to the larger values, maximally up to
$n_2=1$, and the coherent information becomes a non-zero value.

\begin{figure}
\begin{center}
\epsfxsize=6.cm \epsfclipon  \leavevmode  \epsffile{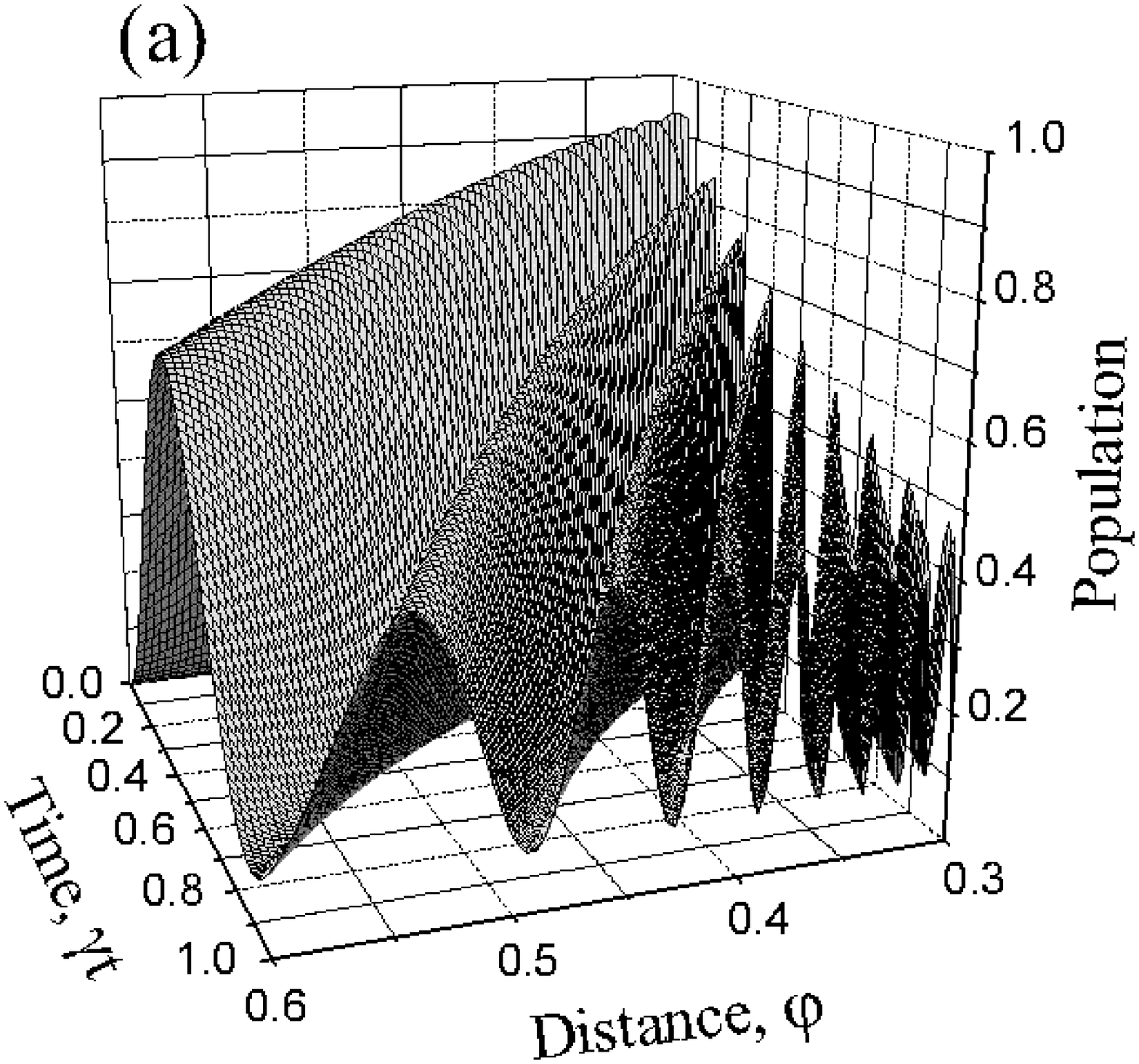}
\epsfxsize=6.cm \epsfclipon  \leavevmode  \epsffile{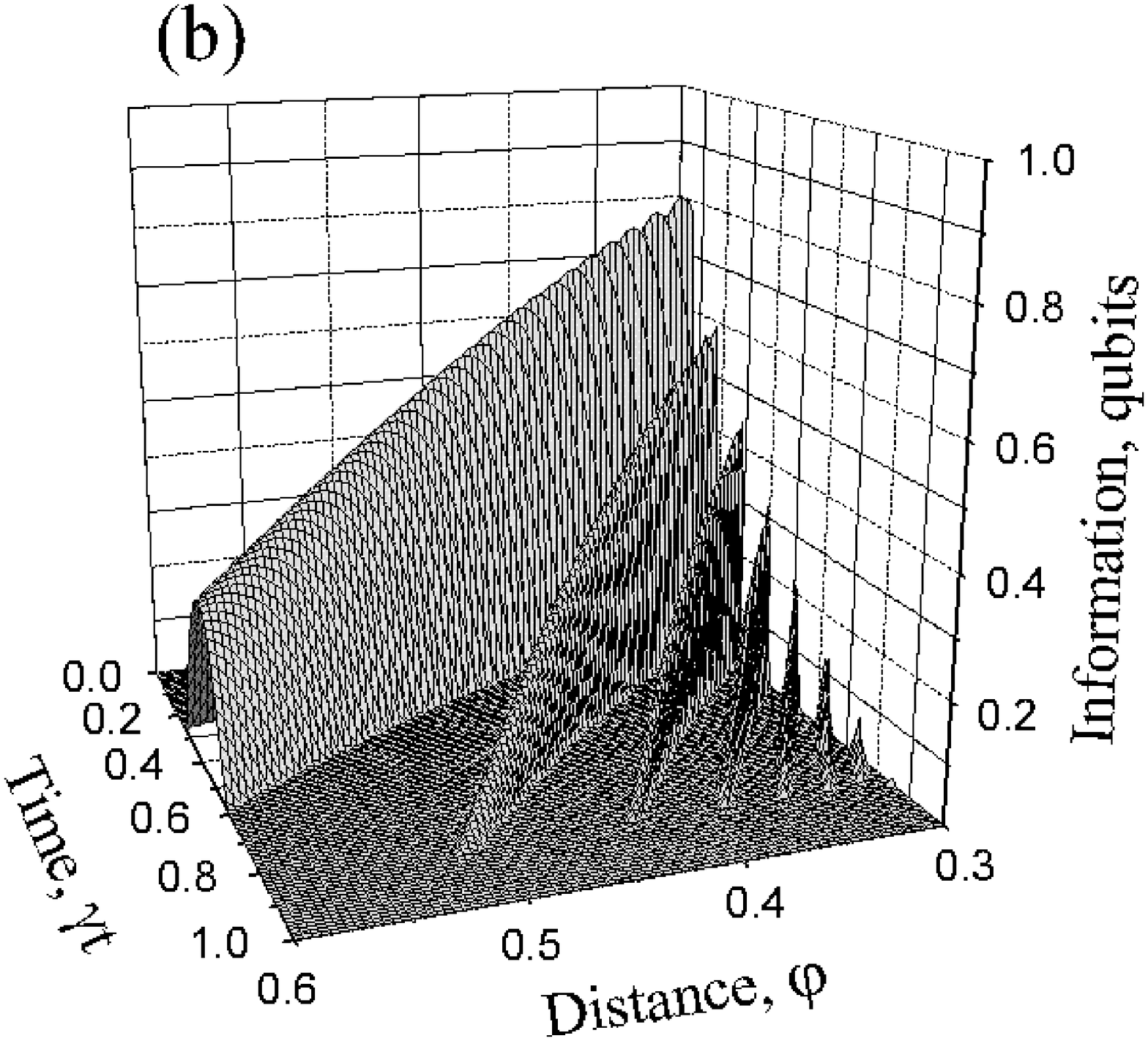}
\end{center}
\caption{Excited state population of the second atom (a) and the
coherent information (b) in a system of two atoms interacting via the
free space field versus time $\gamma t$ and the interatomic distance
$\varphi=\omega_0R/c$ (both are dimensionless). The input density
matrix is diagonal with the ground state matrix element
$\rho_{22}=1/2$.} \label{fig7}
\end{figure}

Functions $n_2(\varphi, \gamma t)$ and $I_c(\varphi,\gamma t)$,
calculated with Eq.\ (\ref{Ican}) are shown in Fig.\ \ref{fig7}.
For the considered geometry, they serve as the universal measures
for a system of two atoms independent of their frequency or dipole
moments.

As can be seen from Fig.\ \ref{fig7}a, the population decreases
rapidly versus time because of the decay of the short-lived Dicke
component. Both the population and the coherent information (Fig.\
\ref{fig7}b) show strong oscillations at smaller interatomic
distances $\varphi$. At $\varphi\to0$ the long-lived Dicke state
yields an essential population even at infinitely long times, but
it does not yield any coherent information after the total decay
of the other short-lived Dicke state.

\section{Conclusions}
\label{sec:conclusions}

In this paper, we have shown that the coherent information concept
can be used effectively to quantify the interaction between two
real quantum systems, which in general case may have essentially
different Hilbert spaces, and to elucidate the role of quantum
coherence specific for the joint system.

For a TLA in a resonant laser field, coherent information in the
system does not increase as the intensity of the external field
increases, unless the external field modifies the relaxation
parameters.

As an example of information transmission between the subsystems
of a whole system, the hydrogen atom was considered. The coherent
information in the atom was shown to transfer from the forbidden
atomic transition to the dipole active transition in an external
electric field, due to coupling through Stark splitting.

For two unitary coupled TLAs, the maximum value $I_c=1$ qubit of
the coherent information was shown to be achieved for a complete
unitary entanglement of two TLAs and $I_c=0$, for any kind of
measuring procedure discussed in subsection \ref{subsec:twoqbm}.

For the information exchange between a TLA and a free-space vacuum
photon field via spontaneous emission, the coherent information
was shown to reach a non-zero value at the threshold point of the
decay exponent $\exp(-\gamma t)$ equal to 1/2, when the
probability of finding no photon is equal to the population of the
lower atomic state. At its maximum, the coherent information can
reach the value of $I_c=1$ qubit.

For the information transfer between two atoms via vacuum field,
when the atoms are located at a distance of the order of their
transition wavelength, the coherent information was shown to be a
non-zero value, only because of the coherent oscillations of the
Dicke states, which originate from the dipole-to-dipole short
distance electrostatic-like $\sim1/R^3$ interaction. In contrast,
the semiclassical information received from the quantum detection
procedure results from the population correlations \cite{lasphys}.

\acknowledgements

This work was partially supported by the programs ``Fundamental
Metrology", ``Physics of Quantum and Wave Phenomena", and
``Nanotechnology" of the Russian Ministry of Science and
Technology. The help of C.~M.~Elliott in preparing the manuscript
is much appreciated.


\begin{table}
\caption{Unitary (atom+field) to (atom+field) transformation
$U_{m\mu,k\alpha}$ for the vacuum initial photon field state, where
indexes $m,k$ stand for atomic quanta and $\mu,\alpha$---for the
number of photons. Long dash symbol stays for the elements not involved
into the calculated terms $S_{kl,\mu\nu}$ (Table \ref{table2}).}
\begin{tabular}{ccccc}
~~~~~~$m\mu$ & \multicolumn{1}{c}{00} & \multicolumn{1}{c}{01}
& \multicolumn{1}{c}{10} & \multicolumn{1}{c}{11} \\
\unitlength1pt
\begin{picture}(-3,0)
\put(0,21){\line(1,-1){24}}
\end{picture}
$k\alpha$  \\ \hline
00& 1 & 0 & 0 & 0 \\
01& --- & --- & --- & --- \\
10& 0 & $\psi_0({\bf k},\lambda)$ & $c_1$ & 0 \\
11& --- & --- & --- & ---
\end{tabular}
\label{table1}
\end{table}

\begin{table}
\caption{Atom-to-field transformation $S_{kl,\mu\nu}$, which defines
$\ket{k}\bra{l}\to\ket{\mu}\bra{\nu}$ superoperator transformation.
Indexes $k,l$ stand for atomic quanta and $\mu,\nu$---for the
number of photons.}
\begin{tabular}{ccccc}
~~~~~~~$\mu\nu$ & \multicolumn{1}{c}{00} & \multicolumn{1}{c}{01}
& \multicolumn{1}{c}{10} & \multicolumn{1}{c}{11} \\
\unitlength1pt
\begin{picture}(-2,0)
\put(0,21){\line(1,-1){24}}
\end{picture}
$kl$  \\ \hline
00& 1 & 0 & 0 & 0 \\
01& 0 & 0 & $\psi_0({\bf k},\lambda)$ & 0 \\
10& 0 & $\psi_0^+({\bf k},\lambda)$ & 0 & 0 \\
11&$|c_1|^2$&0&0&$\psi_0({\bf k},\lambda)\psi_0^+({\bf k}',\lambda')$
\end{tabular}
\label{table2}
\end{table}

\begin{references}
\bibitem{note}
It is worthwhile to note a qualitative difference between these channels in
connection with the physical casuality principle \cite{einstein}:
specific restrictions due to the casuality principle are essential only
for the channels $1\to2$ and $1\to(1+2)$ because of the spatial
locality of the subsystems 1, 2. With this, analysis of a system of two
atoms interacting with the vacuum photon field, given below, could be
complemented by the relativistic retardation.
\bibitem{positivity}
To check the complete positivity \cite{kraus83}, one has to
introduce the operators $\hat s_{kl}\otimes\hat1$.
\bibitem{conjugation}
Complex conjugation is added here to make the expression invariant
in respect to the rotations in the eigen subspaces corresponding
to the degenerate eigenvalues $p_i$. This correction has no effect
for the matrices $\hat\rho_{\rm in}$ with real matrix elements and
nondegenerate spectrum.
\bibitem{TTLA}
In fact, the results of this section are valid not only for two
TLAs, but also for any two quantum systems having Hilbert spaces of
finite dimensions.
\bibitem{addnote}
It is important to emphasize here that there are no restrictions
on the manipulations with the photon field, so that the emitted
photon appears to be a coherent signal.
\end{references}
\end{document}